\documentclass[aps,prl,twocolumn,groupedaddress,showpacs]{revtex4}

\usepackage{graphicx}
\usepackage{dcolumn}
\usepackage{bm}

\begin{document}

\title{Suppression of Spin Frustration due to Orbital Selection}

\author{Hiroaki Onishi and Takashi Hotta}
\affiliation{Advanced Science Research Center,
Japan Atomic Energy Research Institute,
Tokai, Ibaraki 319-1195, Japan}

\date{October 1, 2004}

\begin{abstract}
In order to clarify a crucial role of orbital degree of freedom
in geometrically frustrated systems,
we investigate both ground- and excited-state properties of
the $e_{\rm g}$-orbital degenerate Hubbard model
on two kinds of lattices, ladder and zigzag chain,
by using numerical techniques.
In the ladder, spin correlation extends on the whole system,
while the zigzag chain is decoupled to a double chain and
spin excitation is confined in one side of the double chain
due to the selection of a specific orbital.
We envision a kind of self-organization phenomenon that
the geometrically frustrated multi-orbital system
is spontaneously reduced to a one-orbital model
to suppress the spin frustration.
\end{abstract}


\maketitle


Recently strongly correlated electron systems with
geometrical frustration have attracted much attention
in the research field of condensed-matter physics,
since a subtle balance among competing interactions
leads to a variety of cooperative phenomena
such as exotic superconductivity and novel magnetism.
In particular, the discovery of superconductivity
in layered cobalt oxyhydrate Na$_{0.35}$CoO$_2$$\cdot$1.3H$_2$O
\cite{Takada} has triggered intensive investigations of
superconductivity on the triangular lattice.
Concerning magnetism on geometrically frustrated lattices,
properties of antiferromagnets with the triangle-based structure
have been discussed for a long time \cite{review}.
In low-dimensional systems,
the combined effects of geometrical frustration
and strong quantum fluctuation
cause peculiar behavior in low-energy physics,
as typically observed in the Heisenberg zigzag chain with spin $S$=1/2.
With increasing the strength of frustration,
the ground state is changed
from a critical spin-liquid to a gapped dimer phase
\cite{Majumdar-Ghosh,Tonegawa-zigzag,Okamoto-zigzag,White-zigzag}.
In the dimer phase,
neighboring spins form a valence bond to gain the local magnetic energy,
while the correlation among the valence bonds is weakened
to suppress the effect of spin frustration.

Another important ingredient in actual materials is
$orbital$ degree of freedom,
when electrons partially fill degenerate orbitals,
as frequently observed in $d$- and $f$-electron compounds.
In such a system,
the interplay of spin and orbital degrees of freedom yields
the possibility of orbital ordering.
It is an intriguing issue to clarify the influence of
orbital ordering on magnetic properties
in geometrically frustrated systems.
In fact, significance of $t_{\rm 2g}$-orbital degree of freedom
has been discussed to understand the mechanism of
two phase transitions
in spinel vanadium oxides $A$V$_2$O$_4$ ($A$=Zn, Mg, and Cd)
\cite{Tsunetsugu-AV2O4,Motome-AV2O4,Lee-ZnV2O4}.
It has been proposed that
orbital ordering brings a spatial modulation in the spin exchange
and spin frustration is consequently relaxed.
Similarly, for MgTi$_2$O$_4$,
the formation of a valence-bond crystal due to orbital ordering
has been also suggested \cite{Matteo-MgTi2O4}.

In general, since $d$- and $f$-electron orbitals are spatially
anisotropic, there always exist easy and hard directions
for electron motion.
Then, it is reasonable to expect that the effect of geometrical
frustration would be reduced due to orbital ordering,
depending on the lattice structure and the type of orbital,
in order to arrive at the spin structure to minimize
the influence of spin frustration.
However, the spin structure on such an orbital-ordered background
may be fragile, since the effect of geometrical frustration never
vanishes, unless the lattice distortion is explicitly taken
into account.
It is a highly non-trivial issue whether such an orbital arrangement
actually describes the low-energy physics of
geometrically frustrated systems.
In particular, it is quite important to clarify
how the orbital-arranged background is intrinsically stabilized
through the spin-orbital correlation
even without the electron-lattice coupling.

In this Letter, we investigate both ground- and excited-state
properties of the $e_{\rm g}$-orbital degenerate Hubbard model
on a ladder and a zigzag chain (see Fig.~1)
by exploiting numerical techniques.
In the ladder, ferro-orbital ordering occurs,
but the spin correlation on the rung remains finite
in a paramagnetic (PM) ground state.
On the other hand, the zigzag chain is found to be decoupled
to a double chain in the ground state, while
in the first spin-excited state, we find that
the spin-triplet excitation is localized
just in one side of the double chain.
The robust decoupling nature of the zigzag chain is understood
by a concept that the multi-orbital system with geometrical
frustration is spontaneously reduced to a one-orbital model
to suppress the spin frustration
at the sacrifice of orbital degree of freedom.
This is a kind of self-organizing phenomenon, characteristic of
the systems with a subtle balance among competing interactions.

\begin{figure}[b]
\includegraphics[width=1.0\linewidth]{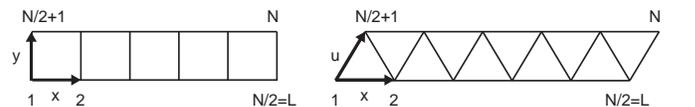}
\caption{
Lattice location and site numbering of $N$-site ladder
and zigzag chain. The length is defined as $L$=$N/2$.
}
\end{figure}


Let us consider doubly degenerate $e_{\rm g}$ orbitals
on the $N$-site ladder and zigzag chain
including one electron per site (quarter filling).
Here the lattices are located in the ($x,y$) plane,
as shown in Fig.~1.
Note that the zigzag chain is composed of equilateral triangles.
The $e_{\rm g}$-orbital degenerate Hubbard model is given by
\begin{eqnarray}
 H &=&
 \sum_{{\bf i},{\bf a},\gamma,\gamma',\sigma}
 t_{\gamma\gamma'}^{\bf a}
 d_{{\bf i}\gamma\sigma}^{\dag} d_{{\bf i}+{\bf a}\gamma'\sigma}
 +U \sum_{{\bf i},\gamma}
 \rho_{{\bf i}\gamma\uparrow} \rho_{{\bf i}\gamma\downarrow}
 \nonumber\\
 &&
 +U' \sum_{{\bf i},\sigma,\sigma'} 
 \rho_{{\bf i}a\sigma} \rho_{{\bf i}b\sigma'}
 +J \sum_{{\bf i},\sigma,\sigma'} 
 d_{{\bf i}a\sigma}^{\dag} d_{{\bf i}b\sigma'}^{\dag}
 d_{{\bf i}a\sigma'} d_{{\bf i}b\sigma}
 \nonumber\\
 &&
 +J' \sum_{{\bf i},\gamma \ne \gamma'} 
 d_{{\bf i}\gamma\uparrow}^{\dag} d_{{\bf i}\gamma\downarrow}^{\dag}
 d_{{\bf i}\gamma'\downarrow} d_{{\bf i}\gamma'\uparrow},
\end{eqnarray}
where $d_{{\bf i}a\sigma}$ ($d_{{\bf i}b\sigma}$)
is the annihilation operator for an electron with spin $\sigma$
in the $d_{3z^2-r^2}$ ($d_{x^2-y^2}$) orbital at site ${\bf i}$,
$\rho_{{\bf i}\gamma\sigma}$=%
$d_{{\bf i}\gamma\sigma}^{\dag}d_{{\bf i}\gamma\sigma}$,
${\bf a}$ is the vector connecting adjacent sites,
and $t_{\gamma,\gamma'}^{\bf a}$ is the hopping amplitude
between adjacent $\gamma$ and $\gamma'$ orbitals
along the ${\bf a}$ direction.
The hopping amplitudes depend on the bond direction as
$t_{aa}^{\bf x}$=$t/4$,
$t_{ab}^{\bf x}$=$t_{ba}^{\bf x}$=$-\sqrt{3}t/4$,
$t_{bb}^{\bf x}$=$3t/4$
for the $x$ direction,
$t_{aa}^{\bf y}$=$t/4$,
$t_{ab}^{\bf y}$=$t_{ba}^{\bf y}$=$\sqrt{3}t/4$,
$t_{bb}^{\bf y}$=$3t/4$
for the $y$ direction, and
$t_{aa}^{\bf u}$=$t/4$,
$t_{ab}^{\bf u}$=$t_{ba}^{\bf u}$=$\sqrt{3}t/8$,
$t_{bb}^{\bf u}$=$3t/16$
for the oblique $u$ direction.
Hereafter, $t$ is taken as the energy unit.
In the interaction terms,
$U$ is the intraorbital Coulomb interaction,
$U'$ the interorbital Coulomb interaction,
$J$ the interorbital exchange interaction,
and $J'$ is the pair-hopping amplitude between different orbitals.
Note that the relation $U$=$U'$+$J$+$J'$ holds
due to the rotation symmetry
in the orbital space and $J'$=$J$ is assumed
due to the reality of the wave function.


In order to analyze the complex model including $both$
the orbital degree of freedom and geometrical frustration,
we employ the finite-system density matrix
renormalization group (DMRG) method
appropriate for quasi-one-dimensional systems
with the open boundary condition \cite{White-DMRG,Liang-DMRG}.
Since one site includes two $e_{\rm g}$ orbitals and
the number of bases is 16 per site,
the size of the superblock Hilbelt space becomes
very large as $m^2$$\times$$16^2$,
where $m$ is the number of states kept for each block.
Thus, we treat each orbital as an effective site to reduce the size
of the superblock Hilbert space to $m^2$$\times$$4^2$.
In the present calculations, we keep $m$ states up to $m$=200
and the truncation error is estimated to be $10^{-5}$ at most.
Since the DMRG calculations have consumed CPU times,
we supplementary use the Lanczos method for $N$=8 to study
ground-state properties with relatively short CPU times.


\begin{figure}[t]
\includegraphics[width=1.0\linewidth]{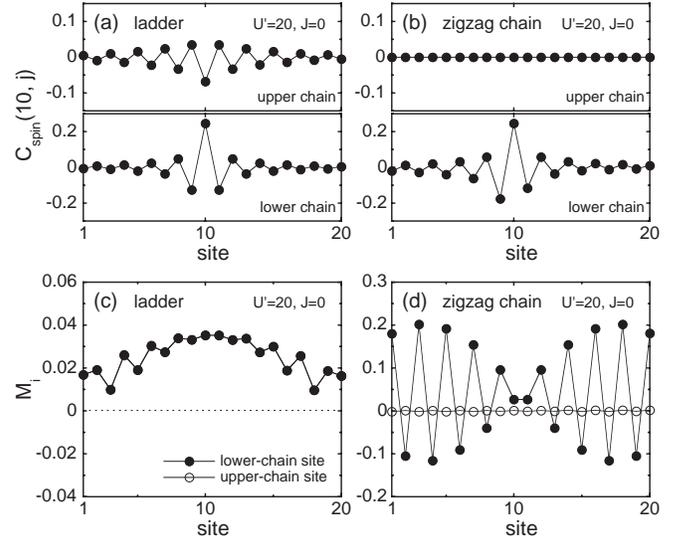}
\caption{
The spin-correlation function measured from the center
of the lower chain for the PM ground state
in (a) the ladder and (b) the zigzag chain.
The local magnetization for the first spin-excited state
in (c) the ladder and (d) the zigzag chain.
}
\end{figure}

First we focus on the spin structure of the PM ground state
at $J$=0, since the zigzag chain is relevant to a geometrically
frustrated antiferromagnet in the spin-singlet PM phase.
The effect of $J$ will be discussed later.
In Figs.~2(a) and (b), we show the DMRG results for $N$=40
of the spin-correlation function
$C_{\rm spin}({\bf i},{\bf j})$=%
$\langle S_{\bf i}^z S_{\bf j}^z \rangle$
with
$S_{\bf i}^z$=$\sum_{\gamma}(\rho_{{\bf i}\gamma\uparrow}$%
$-$$\rho_{{\bf i}\gamma\downarrow})/2$.
Note that a large value of $U'$=20 is used
to consider the strong-coupling region,
but the results do not change qualitatively for small values of $U'$.
As shown in Fig.~2(a),
we observe a simple N\'eel structure in the ladder.
On the other hand, in the zigzag chain,
there exists an antiferromagnetic (AFM) correlation
between intra-chain sites in each of lower and upper chains,
while the spin correlation between inter-chain sites is much weak,
as shown in Fig.~1(b).
Namely, the zigzag chain seems to be decoupled to a double chain
in terms of the spin structure.

In order to clarify the characteristics of the spin structure
in the excited state, we investigate the local magnetization
$M_{\bf i}$=$\langle S_{\bf i}^z \rangle$
for the lowest-energy state with $S_{\rm tot}^z$=1,
i.e., the first spin-excited state,
where $S_{\rm tot}^z$ is the $z$ component of the total spin.
In the ladder, the total moment of $S_{\rm tot}^z$=1 is distributed
to the whole system and there is no significant structure,
as shown in Fig.~2(c).
On the other hand, the situation is drastically changed
in the zigzag chain.
As shown in Fig.~2(d), the total moment of
$S_{\rm tot}^z$=1 is confined in the lower chain
and it forms a sinusoidal shape with a node,
while nothing in the upper chain.
Note that the sinusoidal shape of the local magnetization is
a characteristic feature of the $S$=1/2 AFM Heisenberg chain
with edges at low temperatures \cite{Laukamp-edge,Nishino-edge}.
Thus, the double-chain nature in the spin structure remains robust
even for the spin-excited state.


Let us now consider the orbital arrangement to understand the mechanism
of the appearance of the spin structures.
In order to determine the orbital arrangement,
we usually measure orbital correlations,
but due care should be paid to the definition.
By using an angle $\theta_{\bf i}$ to characterize the orbital shape
at each site, we introduce new phase-dressed operators as
\cite{Hotta-berryphase}
\begin{equation}
 \left\{
 \begin{array}{l}
 \tilde{d}_{{\bf i}a\sigma}=
 e^{i\theta_{\bf i}/2}[\cos(\theta_{\bf i}/2)d_{{\bf i}a\sigma}+
 \sin(\theta_{\bf i}/2)d_{{\bf i}b\sigma}],
 \\
 \tilde{d}_{{\bf i}b\sigma}=
 e^{i\theta_{\bf i}/2}[-\sin(\theta_{\bf i}/2)d_{{\bf i}a\sigma}+
 \cos(\theta_{\bf i}/2)d_{{\bf i}b\sigma}].
 \end{array}
 \right.
\end{equation}
The optimal set of $\{\theta_{\bf i}\}$ is determined
so as to maximize the Fourier transform of
the orbital-correlation function,
\begin{equation}
 T({\bf q})=
 (1/N^{2})\sum_{{\bf i},{\bf j}}
 \langle \tilde{T}_{\bf i}^{z}\tilde{T}_{\bf j}^{z} \rangle
 e^{i{\bf q}\cdot({\bf i}-{\bf j})},
\end{equation}
with $\tilde{T}_{\bf i}^z$=$\sum_{\sigma}
(\tilde{d}_{{\bf i}a\sigma}^{\dag}\tilde{d}_{{\bf i}a\sigma}$$-$%
$\tilde{d}_{{\bf i}b\sigma}^{\dag}\tilde{d}_{{\bf i}b\sigma})/2$.

\begin{figure}[t]
\includegraphics[width=1.0\linewidth]{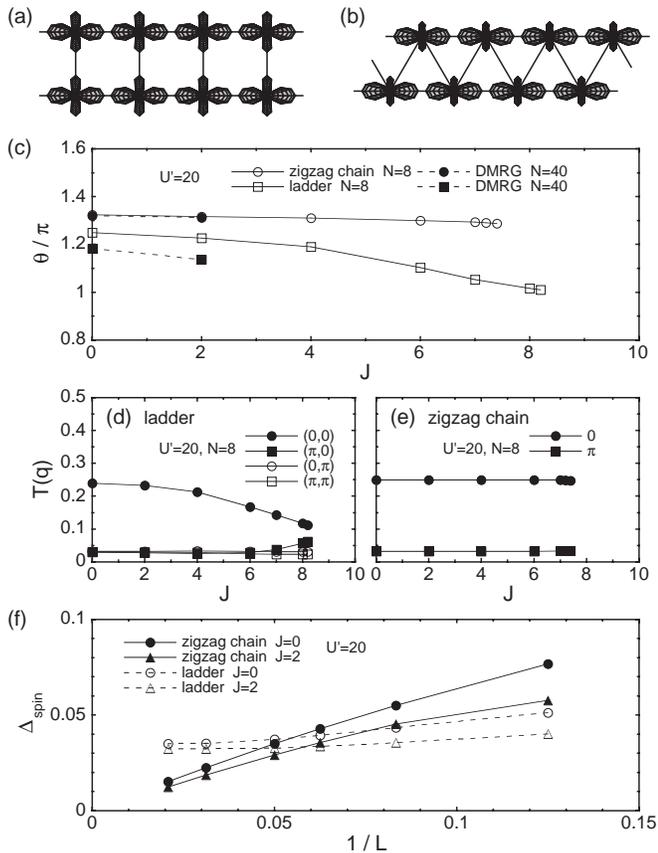}
\caption{
Optimal orbital arrangement in (a) the ladder and (b) the zigzag chain.
(c) The angle representing the orbital structure
in the PM ground state.
The maximum value of $T({\bf q})$ at each wave vector
is shown for (d) the zigzag chain and (e) the ladder.
Note that in the zigzag chain,
the wave vector is defined along the one-dimensional zigzag path.
(f) The spin gap as a function of $1/L$ with $L$=$N/2$.
}
\end{figure}

In the ladder, we observe ferro-orbital (FO) ordering,
characterized by $\theta_{\bf i}/\pi$$\sim$1.18
in the ground state, as shown in Fig.~3(a).
In the first spin-excited state, the FO structure also appears,
but the angle characterizing the orbital shape is slightly
changed as $\theta_{\bf i}/\pi$$\sim$1.20
to further extend to the leg direction.
On the other hand, in the zigzag chain,
we observe that $both$ in the ground and
first spin-excited states, $T({\bf q})$ becomes maximum
at ${\bf q}$=0 with $\theta_{\bf i}/\pi$$\sim$$1.32$,
indicating a 3$x^2$$-$$r^2$ orbital at each site,
as shown in Fig.~3(b).
It is emphasized here that the orbital arrangement is unchanged even
in the spin-excited state.
We envisage that the orbital degree of freedom spontaneously
becomes ``dead'' in low-energy states
to suppress the effect of spin frustration.


Consider now the effect of $J$ on the spin-orbital structure
in the PM ground state.
It is useful to show the results for 8-site system with
the periodic boundary condition obtained by the Lanczos method
in addition to a few DMRG results for $N$=40.
In Fig.~3(c), we depict the angle $\theta$ representing
the orbital structure, where $\theta$ is the optimal angle
taking the same value at all sites.
In the ladder, $\theta$ decreases with the increase of $J$,
indicating that the orbital shape is changed
to further extend to the rung direction.
Note that $\theta$ for $N$=40 is smaller than that for $N$=8.
The orbital shape is sensitive to the length of the leg,
since $\theta$ is determined to optimize the electron motion
both in the leg and rung directions.
When we measure the maximum value of $T({\bf q})$ at each
wave vector, the amount of the FO correlation $T(0,0)$
is reduced due to the effect of $J$, as shown in Fig.~3(d).
Note that $T(\pi,0)$ grows with the increase of $J$,
while the spin correlation of ${\bf q}$=($0,\pi$)
is found to develop at the same time (not shown here).
The development of this type of the spin-orbital correlation
indicates that the orbital degree of freedom is still active,
since electrons occupy $b$ orbitals to some extent
even in the $a$-orbital ordered state.

In the zigzag chain, however, no remarkable change is
found in the angle $\theta$ due to the effect of $J$,
as shown in Fig.~3(c).
In sharp contrast to the case of the ladder,
the finite-size effect on the orbital shape is very small,
since the orbital polarization in this case occurs
due to the geometrical reason.
As shown in Fig.~3(e), irrespective of the value of $J$,
$T(0)$ takes 1/4, which is the possible maximum value,
indicating the perfect orbital polarization.
All these facts suggest that {\it the PM phase of the zigzag chain
is described by a one-orbital model},
since one of the doubly degenerate orbitals is selectively occupied.
Thus, the double-chain nature of the spin structure remains robust
even for the inclusion of $J$.


Let us discuss the anisotropy in the spin-exchange interactions.
For the zigzag chain described by the Hubbard model
composed of 3$x^2$$-$$r^2$ orbital,
since the orbital shape extends along the double chain,
not along the zigzag path,
the AFM exchange interaction along the $u$ direction $J_1$
should be much weaker than that along the $x$ direction $J_2$,
estimated as $J_1/J_2$=$1/64^2$ \cite{note1}.
Namely, the spin correlation on the zigzag path is
reduced due to the spatial anisotropy of 3$x^2$$-$$r^2$ orbital.
Thus, the zigzag chain is effectively regarded as
a double-chain system of the $S$=1/2 AFM Heisenberg chain.
Concerning the low-energy physics,
it is intuitively expected that
the spin gap should be extremely suppressed,
since the spin gap decreases exponentially with the increase of $J_2/J_1$
in the gapped dimer phase in the zigzag spin chain \cite{White-zigzag}.

On the other hand, in the ladder with the FO structure as shown
in Fig.~3(a), the orbital shape extends to the rung direction
as well as to the leg direction.
In this case, 
$J_{\rm rung}/J_{\rm leg}$ is estimated to be 0.26 \cite{note2},
much larger than $J_1/J_2$=$1/64^2$ in the zigzag chain,
where $J_{\rm leg}$ and $J_{\rm rung}$ are the AFM exchange interactions
along the leg and rung directions, respectively.
Thus, the spin correlation on the rung remains finite,
leading to the simple N\'eel structure,
and the spin excitation in the ladder is expected to be gapful
similar to the spin ladder \cite{Barnes-ladder,Greven-ladder}.
Note that for the spin ladder of $J_{\rm rung}/J_{\rm leg}$=0.26,
the spin gap is estimated to be 0.02 in the present energy unit.

Let us show our numerical results on the spin gap.
We investigate the system-size dependence of the spin gap
$\Delta_{\rm spin}$=$E(N/2+1,N/2-1)$$-$$E(N/2,N/2)$ up to $N$=96,
where $E(N_\uparrow,N_\downarrow)$ denotes the lowest energy
in the subspace with $N_\uparrow$ up- and
$N_\downarrow$ down-spin electrons.
As shown in Fig.~3(f), the spin gap of the zigzag chain
actually converges to almost zero in the thermodynamic limit,
as expected by analogy with the zigzag spin chain.
Note that the spin gap becomes small with the increase of $J$
for each system size.
On the other hand, a finite spin gap is observed in the ladder.
Note, however, that the value of the spin gap is somewhat larger
than that expected by analogy with the spin ladder,
since the multi-orbital ladder may not
be regarded as a simple spin ladder.

\begin{figure}[t]
\includegraphics[width=1.0\linewidth]{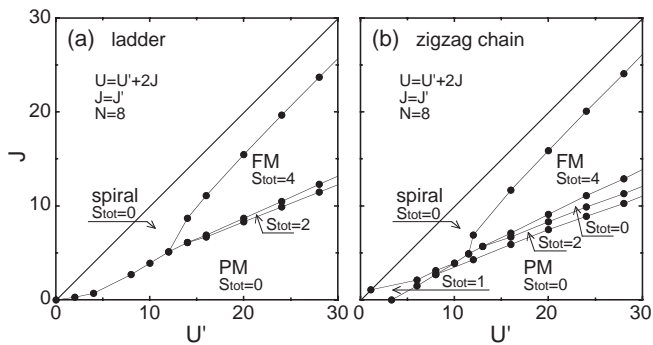}
\caption{
The ground-state phase diagram for $N$=8 of
(a) the zigzag chain and (b) the ladder.
}
\end{figure}


Finally, let us briefly discuss the characteristics of
the phase diagram beyond the region of small $J$.
Here we show the Lanczos results for $N$=8, but
qualitative points are believed to be grasped in these calculations.
In Figs.~4(a) and (b),
we show the ground-state phase diagrams in the ($U',J$) plane
for the ladder and the zigzag chain, respectively.
Note that the region of $J$$>$$U'$ is ignored, since it is unphysical.
Each phase is characterized by the total spin $S_{\rm tot}$,
which is determined by comparing the lowest energies
in the subspaces of $S_{\rm tot}^z$=0$\sim$4.
Roughly speaking, there appear three types of ground-state phases
both in the zigzag chain and the ladder.
In the region of small $J$, as discussed intensively in this paper,
the ground state is the PM phase
characterized by $S_{\rm tot}$=0.
In the region of large $U'$ and $J$,
a maximally spin-polarized ferromagnetic (FM) phase occurs,
where the lowest energies in the subspaces of
$S_{\rm tot}^z$=0$\sim$4 are all degenerate.
In the region next to the $U'$=$J$ line,
the ground state is characterized by $S_{\rm tot}$=0 again,
where we observe a spin-spiral state,
which is considered to be an FM state
in the thermodynamic limit \cite{Arita-spiral}.
In addition to these three types of ground-state phases,
we find narrow regions with $S_{\rm tot}$=0, 1, and/or 2.
The existence of such a partial FM state has been discussed
in a single-orbital Hubbard model
on the zigzag chain \cite{Nakano-PF}.
It is an interesting future problem to clarify the feature of the
partial FM state in the multi-orbital system.


In summary, we have discussed both ground- and excited-state properties
of the $e_{\rm g}$-orbital degenerate Hubbard model
on the ladder and the zigzag chain.
It has been found that the zigzag chain is reduced to
a decoupled double-chain spin system
due to the selection of a specific orbital.
It is considered as a general feature of geometrically frustrated
multi-orbital systems
that the orbital selection spontaneously occurs
so as to suppress the effect of spin frustration.


One of the authors (T.H.) is supported by Grant-in-Aids for
Scientific Research of Japan Society for the Promotion of Science
and for Scientific Research in Priority Area ``Skutterudites''
of the Ministry of Education, Culture, Sports,
Science, and Technology of Japan.


\end{document}